\documentstyle[tighten,preprint,eqsecnum,aps,prd]{revtex}
\begin{document}
\draft

\hyphenation{
mani-fold
mani-folds
Schwarz-schild
}


\def\casehalf{{\case{1}{2}}}
\def\sfR{{\sf R}}
\def\sfF{{\sf F}}

\def\lambdahat{{\hat\lambda}}

\def\bm{{\bf m}}

\def\bp{{\bf p}}

\def\bh{{\bf h}}


\preprint{\vbox{\baselineskip=12pt
\rightline{PP97--52}
\rightline{WISC--MILW--96--TH--26}
\rightline{gr-qc/9610071}}}
\title{Hamiltonian thermodynamics of a Lovelock black hole}
\author{Jorma Louko\footnote{On leave of absence from
Department of Physics, University of Helsinki.
Electronic address:
louko@wam.umd.edu}
and
Jonathan Z. Simon\footnote{Present address:
Institute for Systems Research,
University of Maryland,
College Park, Maryland 20742, USA\null.
Electronic address:
jzsimon@isr.umd.edu}}
\address{
Department of Physics,
University of Maryland,
College Park,
Maryland 20742--4111,
USA}
\author{Stephen N. Winters-Hilt\footnote{Electronic address:
winters@csd.uwm.edu}}
\address{
Department of Physics,
University of
Wisconsin--Milwaukee,
\\
P.O.\ Box 413,
Milwaukee, Wisconsin 53201, USA}
\date{October 1996}
\maketitle
\begin{abstract}%
We consider the Hamiltonian dynamics and thermodynamics of spherically
symmetric spacetimes within a one-parameter family of five-dimensional
Lovelock theories. We adopt boundary conditions that make every classical
solution part of a black hole exterior,
with the spacelike hypersurfaces extending from the horizon bifurcation
three-sphere to a timelike boundary with fixed intrinsic metric. The
constraints are simplified by a Kucha\v{r}-type canonical transformation, and
the theory is reduced to its true dynamical degrees of freedom. After
quantization, the trace of the analytically continued Lorentzian time
evolution operator is interpreted as the partition function of a
thermodynamical canonical ensemble. Whenever the partition function is
dominated by a Euclidean black hole solution, the entropy is given by the
Lovelock analogue of the Bekenstein-Hawking entropy; in particular, in the
low temperature limit the system exhibits a dominant classical solution
that has no counterpart in Einstein's theory. The asymptotically flat space
limit of the partition function does not exist. The results indicate
qualitative robustness of the thermodynamics of five-dimensional Einstein
theory upon the addition of a nontrivial Lovelock term.
\end{abstract}

\narrowtext

\section{Introduction}
\label{sec:intro}

A gravitational theory whose Lagrangian density consists of multiples of
lower-dimensional Euler densities has the property that the field
equations are second order in the metric \cite{lovelock1,lovelock2}. These
theories, known as Lovelock theories, include Einstein's theory with a
cosmological constant in all dimensions greater than two, and in five or
more dimensions they provide genuine curvature squared generalizations of
Einstein's theory. Among all curvature squared generalizations of
Einstein's theory, Lovelock theories therefore have a special status in that
they preserve the number of degrees of freedom: a generic curvature squared
action yields field equations that are fourth order in the metric,
containing thus more degrees of freedom than Einstein's theory.
This has generated wide interest in Lovelock theories, especially in the
contexts of cosmology and black hole physics
\cite{boul-des,jwheeler,wiltshire1,myers,teit-zan-love1,teit-zan-love2,%
wiltshire2,whitt,myers-simon,myers-simon2,melmed-thesis,poisson1,poisson2,%
love-cosmo,jacobson-myers-prl,JGM1,JGM2}.

The purpose of the present paper is to analyze the classical and
quantum dynamics of spherically symmetric Lovelock gravity by the
Hamiltonian methods recently developed by Kucha\v{r} \cite{kuchar1}. These
methods have previously been applied to spherically symmetric
Einstein(-Maxwell) gravity in four dimensions
\cite{kuchar1,LW2,oliveira,lau1,lou-win}, vacuum dilatonic gravity in two
dimensions \cite{lau1,varadarajan,BLPP}, and to related
systems \cite{romano-torre,KRV}; for related discussion, see
Refs.\cite{thiemann1,thiemann2,thiemann3,thiemann4,kuns1,kuns2,kuns3}.
At the classical level, we wish to find a canonical transformation that
introduces the mass parameter of the spacetime as a new canonical variable,
use this transformation to simplify the constraints, and reduce the theory to
its true dynamical degrees of freedom. At the quantum level, we wish to
derive from the quantum theory a partition function that describes the
equilibrium thermodynamics of a Lovelock black hole in the canonical
ensemble.

The issues of prime interest are twofold. First, although
Lovelock theories have the same set of canonical variables as Einstein's
theory, the Lovelock Hamiltonian is, in general, a multivalued function of
the canonical variables \cite{teit-zan-love1,teit-zan-love2}. One anticipates
that this multivaluedness may introduce additional features in the canonical
formulation and Hamiltonian reduction, even though the Lovelock analogue of
Birkhoff's theorem \cite{wiltshire1} strongly suggests that the local
considerations should differ little from those in Einstein's theory. Second,
certain Lovelock black holes have thermodynamical properties that differ
qualitatively from those of Einstein black holes; in particular, a Lovelock
black hole can be stable against Hawking evaporation in asymptotically flat
space \cite{myers-simon}. This leads one to ask whether Lovelock
theories might admit quantum thermodynamical ensembles with boundary
conditions that do not give rise to well-defined ensembles in Einstein's
theory.

The number of possible Lovelock terms in the action increases with the
dimension of the spacetime, and different choices for the coefficients yield
qualitatively different theories. In this paper we shall aim for concreteness
at the expense of generality. We concentrate on a  specific one-parameter
family of Lovelock theories exhibiting both a multivalued Hamiltonian and
asymptotically flat black hole solutions that are stable against Hawking
evaporation.

We take the only bulk contributions to the action to be the
Einstein-Hilbert term and the four-dimensional Euler density. In
$D$ spacetime dimensions, the action then reads \cite{myers-simon}
\begin{eqnarray}
S
&=&
{1 \over 2\kappa } \int d^D x \, \sqrt{-g}
\left[ R
+ {\lambda \over 2}
\left(
R_{abcd} R^{abcd} - 4 R_{ab} R^{ab} + R^2
\right)
\right]
\nonumber
\\
\noalign{\vskip 1\jot}
&&+ \ \ \hbox{(boundary terms)}
\ \ ,
\label{love-action-gen}
\end{eqnarray}
where $\kappa$ is the $D$-dimensional gravitational constant and $\lambda$
is the single Lovelock parameter.\footnote{We have set $c=\hbar=1$: the
gravitational constant $\kappa$ has the dimension of
$\hbox{(length)}^{D-2}$, and the Lovelock parameter $\lambda$ has the
dimension of $\hbox{(length)}^2$.}
For $D\ge5$, the four-dimensional Euler density contributes to the equations
of motion, and we obtain a one-parameter family of generalizations of
Einstein's theory. In these theories, asymptotically flat black hole
solutions that are stable against Hawking evaporation occur only when $D=5$
and $\lambda>0$ \cite{myers-simon}, and we shall therefore concentrate on this
case.  For the interest of comparison, we shall also include the
limiting case of five-dimensional Einstein theory, $D=5$ and $\lambda=0$.

We shall formulate the spherically symmetric Hamiltonian theory with
thermodynamically motivated boundary conditions similar to those introduced in
Ref.\ \cite{LW2}. On a classical solution, one end of the spacelike
hypersurfaces will be at the bifurcation three-sphere of a nondegenerate
Killing horizon, and the other end will be on a timelike hypersurface in
an exterior region of the spacetime. We shall refer to the two ends
respectively as the ``left" end and the ``right" end, as motivated by the
Penrose diagram in which our classical solutions are embedded in the
right-hand-side asymptotically flat region \cite{myers-simon}: on a solution,
one can think of the left end as the inner one and the right end as the outer
one. At the left end we fix the rate at which the spacelike hypersurfaces are
boosted with respect to the coordinate time, and at the right end we fix the
intrinsic metric on the timelike hypersurface.

For $\lambda>0$, the super-Hamiltonian $H$ turns out to be a multivalued
function of the canonical variables. However, our boundary conditions are
sufficient to uniquely determine $H$ near the left end of the spacelike
hypersurfaces, and this solution for $H$ can then be uniquely extended to the
full spacelike hypersurfaces by continuity. Our boundary conditions at the
horizon thus eliminate the difficulties due to the multivaluedness of the
super-Hamiltonian.

We shall find that the theory admits a natural generalization of the
canonical transformation of Refs.\ \cite{kuchar1,LW2}. The constraints
become exceedingly simple, and a Hamiltonian reduction leads again to a
single canonical pair of unconstrained degrees of freedom. On a classical
solution, one member of the pair is the mass parameter, and its conjugate
momentum is the difference of the Killing times at the left and right
ends of the spacelike hypersurfaces.

After taking the curvature radius at the right end of the hypersurfaces to be
time-independent, we quantize the reduced theory by Hamiltonian methods.
Following Ref.\
\cite{LW2}, we analytically continue the time evolution operator to
imaginary time and take the trace, and interpret the resulting object as
the partition function of a thermodynamical canonical ensemble. This
ensemble describes black hole spacetimes in a spherical ``box" whose size and
boundary temperature are fixed.

In the special case of Einstein's theory, $\lambda=0$, we find that the
thermodynamical properties of the system are highly similar to
those of the corresponding system in four dimensions \cite{LW2,york1,WYprl}.
For high boundary temperatures, the partition function is dominated
by a black hole that fills most of the box. For low boundary temperatures,
on the other hand, there is no dominant classical solution. One can argue
that the behavior of the partition function suggests a topological phase
transition from a black hole to ``hot flat space" \cite{york1,WYprl}.

For $\lambda>0$, the partition function displays several qualitatively
different regions depending on the relative magnitudes of the box, the
temperature, and~$\lambda$. In the high temperature limit, with the other two
parameters fixed, the partition function is again dominated by a classical
black hole solution that fills most of the box. In the low temperature limit,
with the other two parameters fixed, the partition function is now {\em
also\/} dominated by a black hole solution: this black hole is small compared
with the box, and it has no analogue in Einstein's theory. However, if
$\lambda$ is small compared with the size of the box, the existence of the new
dominating solution at low temperatures only has a minor effect on
the behavior of the partition function. In this sense, we can say that the
qualitative thermodynamical behavior of the pure Einstein system is stable
against the addition of the Lovelock parameter.

When the size of the box is taken to infinity, the partition function does not
have a well-defined limit, neither for $\lambda=0$ nor for $\lambda>0$. For
$\lambda=0$ this is not surprising: just as in four dimensions, it reflects
the fact that a Schwarzschild hole in asymptotically flat space is not stable
against Hawking evaporation \cite{york1,WYprl}. For $\lambda>0$, on the
other hand, the theory does admit asymptotically flat black hole solutions
that are stable against Hawking evaporation \cite{myers-simon}, and one might
therefore have expected the infinite box limit to exist. The reason why this
is not the case becomes apparent when one tries to repeat our analysis
with boundary conditions that replace the right-hand-side timelike boundary
by an asymptotically flat infinity. The classical reduction and the
construction of a quantum theory proceed without difficulty, but the
effective Euclidean action of the system turns out to be unbounded below, and
the formal integral expression for the partition function is divergent. The
effective action has a {\em local\/} minimum, corresponding to the black hole
that is stable against Hawking evaporation \cite{whitingCQG}, but this is not
sufficient to ensure the existence of the full canonical ensemble. Another
system with locally thermodynamically stable classical solutions but no
well-defined canonical ensemble is four-dimensional Einstein-Maxwell theory
with fixed charge in asymptotically flat space \cite{lou-win}.

The rest of the paper is as follows. In section \ref{sec:metric} we
introduce the notation and present the Hamiltonian formulation in the
metric variables. In section \ref{sec:transformation} we perform a canonical
transformation to Kucha\v{r}-type variables and reduce the theory to its
unconstrained Hamiltonian degrees of freedom. The reduced theory is quantized
and the partition function constructed in section~\ref{sec:quantization},
and the thermodynamics is analyzed in section~\ref{sec:thermodynamics}.
Section \ref{sec:discussion} presents a brief summary and discussion.
Appendix \ref{app:love-hole} reviews briefly the classical solutions to the
theory. The classical and quantum mechanical analyses under asymptotically
flat boundary conditions are outlined in appendix~\ref{app:as-flat}\null.

For the rest of paper, we shall use Planckian units that have been tailored
for numerical convenience: we set $c=\hbar=1$ and $\kappa = 6\pi^2$. All
dimensionful constants have thus disappeared.

\section{Metric formulation}
\label{sec:metric}

In this section we introduce the model and present the Hamiltonian
formulation in the metric variables.

We begin with the general five-dimensional spherically symmetric
Arnowitt-Deser-Misner (ADM) metric,
\begin{equation}
ds^2 = - N^2 dt^2 + \Lambda^2 {(dr + N^r dt)}^2 +R^2 d\Omega_3^2
\ \ .
\label{5-metric}
\end{equation}
Here $d\Omega_3^2$ is the metric on the unit three-sphere, and $N$,
$N^r$, $\Lambda$, and $R$ depend on the coordinates $t$ and $r$ only.
The coordinate $r$ has the range $0\le r \le 1$; this is convenient in view
of our boundary conditions, which will make the radial proper
distance on the constant $t$ hypersurfaces finite.
Unless otherwise stated, we assume
both the spatial metric and the spacetime metric to be nondegenerate.
In particular, we take $\Lambda$, $R$, and $N$ to be positive.

Inserting the metric (\ref{5-metric}) in the Lovelock
action (\ref{love-action-gen}) with
$D=5$, integrating over the three-sphere, and dropping a total derivative,
we recover the action
\begin{equation}
S^L_\Sigma = \int dt \int_0^1 dr \, {\cal L}
\ \ ,
\label{S-lag-sigma}
\end{equation}
where
\begin{eqnarray}
{\cal L} &=& - {\left[{\dot{\Lambda}} - {(N^r\Lambda)}'\right]
({\dot{R}} - N^r R') \over N}
\left\{ R^2 + \lambdahat
\left[
1 - {\left( {R'\over\Lambda}\right)}^2 +
{ {({\dot{R}} - N^r R')}^2 \over 3
N^2 } \right]
\right\}
\nonumber
\\
&&
- { {({\dot{R}} - N^r R')}^2 \over N }
\left[ \Lambda R - \lambdahat {\left( {R'\over\Lambda}\right)}' \right]
\nonumber
\\
&&
+ N \Lambda R
\left[
1 - {\left( {R'\over\Lambda}\right)}^2 \right]
-
N {\left( {R'\over\Lambda}\right)}'
\left\{ R^2 + \lambdahat
\left[
1 - {\left( {R'\over\Lambda}\right)}^2 \right]
\right\}
\ \ .
\label{lagrangian}
\end{eqnarray}
The overdot and the prime denote respectively $\partial/(\partial t)$ and
$\partial/(\partial r)$. We have written $\lambda = \casehalf \lambdahat$,
conforming to the notation of Ref.\ \cite{myers-simon}. We have verified
that the Lagrangian equations of motion obtained from local variations of
$S^L_\Sigma$ (\ref{S-lag-sigma}) are equivalent to the full spherically
symmetric Lovelock equations \cite{boul-des,jwheeler,wiltshire1} derived from
the action~(\ref{love-action-gen}). The reduction of the action by spherical
symmetry is therefore consistent with the equations of motion, and we can
take $S^L_\Sigma$ (\ref{S-lag-sigma}) as the starting point of the dynamical
analysis.  We shall address the boundary conditions and boundary
terms within the Hamiltonian formulation below.

For the reasons discussed in section~\ref{sec:intro}, we take
$\lambdahat\ge0$. For presentational simplicity, we shall assume
$\lambdahat>0$ until explicitly stated otherwise in
section~\ref{sec:thermodynamics}. In the limiting case of five-dimensional
Einstein gravity, $\lambdahat=0$, the analysis would proceed in an entirely
analogous manner, with the obvious technical simplifications.

The Hamiltonian form of the action (\ref{S-lag-sigma}) is
\begin{equation}
S_\Sigma = \int dt \int_0^1 dr
\left(
P_\Lambda {\dot{\Lambda}}
+
P_R {\dot{R}}
-
N H
-
N^r H_r
\right)
\ \ ,
\label{S-sigma}
\end{equation}
where the super-Hamiltonian constraint and the supermomentum constraint are
given respectively by
\begin{mathletters}
\begin{eqnarray}
H &=&
y
\left\{
P_R + y
\left[
\Lambda R - \lambdahat {\left( {R'\over\Lambda}\right)}'
\right]
\right\}
\nonumber
\\
&&
- \Lambda R
\left[
1 - {\left( {R'\over\Lambda}\right)}^2 \right]
+
{\left( {R'\over\Lambda}\right)}'
\left\{ R^2 + \lambdahat
\left[
1 - {\left( {R'\over\Lambda}\right)}^2 \right]
\right\}
\ \ ,
\\
H_r &=& R' P_R - \Lambda P_\Lambda'
\label{supermom}
\ \ .
\end{eqnarray}
\end{mathletters}%
The quantity $y$ is determined in terms of the canonical variables by
the cubic equation
\begin{equation}
0=
\case{1}{3} \lambdahat y^3
+ y
\left\{ R^2 + \lambdahat
\left[
1 - {\left( {R'\over\Lambda}\right)}^2 \right]
\right\}
+ P_\Lambda
\ \ .
\label{y-cubic}
\end{equation}
Note that the form of the supermomentum (\ref{supermom}) is completely
determined by the fact that it must generate spatial diffeomorphisms
in all the canonical variables, together with the observation that $R$ and
$P_\Lambda$ are spatial scalars whereas $\Lambda$ and
$P_R$ are spatial densities \cite{kuchar1}.


Depending on the values of $\Lambda$, $R$, and $P_\Lambda$, the cubic
(\ref{y-cubic}) can have up to three real solutions for~$y$. The
super-Hamiltonian is therefore a potentially multivalued function of the
canonical variables. Such multivaluedness arises generically in Lovelock
theories, owing to the presence of kinetic terms higher than quadratic in the
velocities in the Lagrangian density \cite{teit-zan-love1,teit-zan-love2};
in our case, the highest kinetic terms in
(\ref{lagrangian}) are quartic in the velocities. We shall address this
phenomenon in more detail below. The geometrical meaning of $y$ is revealed
by observing that the equation of motion obtained by varying $S_\Sigma$
(\ref{S-sigma}) with respect to $P_R$ reads
\begin{equation}
{\dot R} = Ny + N^r R'
\ \ .
\label{y-true}
\end{equation}
On a classical solution, $y$ is therefore uniquely determined by the
embedding of the spacelike hypersurface in the spacetime. Conversely,
when multiple real solutions to (\ref{y-cubic}) exist, it can be verified
that they generically lead to different spacetimes.

Let us turn to the boundary conditions. {}From the Lovelock generalization of
Birkhoff's theorem \cite{wiltshire1} it follows that the local properties of
the classical solutions are completely characterized by a discrete binary
parameter and a continuous, mass-like parameter. The general solution is
shown in curvature coordinates in appendix~\ref{app:love-hole}\null. We wish
to concentrate on the black hole solutions, whose global structure is similar
to that of  Kruskal manifold \cite{myers-simon}.  We further wish to attach
the left end of our spacelike hypersurfaces to the bifurcation three-sphere,
and to prescribe there the rate at which the hypersurfaces are boosted with
respect to our coordinate time. The right end of the hypersurfaces will then
be in the right-hand-side exterior region, and we wish to prescribe the metric
on the timelike hypersurface that this end traces. We must now specify
boundary conditions and boundary terms that achieve this.

Consider first the left end of the hypersurfaces. Following the analogous
treatment in Refs.\ \cite{LW2,lau1,lou-win,BLPP}, we adopt at $r\to0$
the falloff
\begin{mathletters}
\label{s-r}
\begin{eqnarray}
\Lambda (t,r) &=& \Lambda_0(t) + O(r^2)
\ \ ,
\label{s-r-Lambda}
\\
R(t,r) &=& R_0(t) + R_2(t) r^2 + O(r^4)
\ \ ,
\label{s-r-R}
\\
P_{\Lambda}(t,r) &=& O(r^3)
\ \ ,
\label{s-r-PLambda}
\\
P_{R}(t,r) &=& O(r)
\ \ ,
\label{s-r-PR}
\\
N(t,r) &=& N_1(t)r + O(r^3)
\ \ ,
\label{s-r-N}
\\
N^r(t,r) &=& N^r_1(t)r + O(r^3)
\ \ ,
\label{s-r-Nr}
\end{eqnarray}
\end{mathletters}%
where $\Lambda_0$ and $R_0$ are positive, and $N_1\ge0$.
$O(r^n)$~stands for a term that is bounded at $r\to0$ by a constant
times~$r^n$, and whose derivatives fall off accordingly.
As in Refs.\ \cite{LW2,lau1,lou-win,BLPP}, these conditions guarantee that the
classical solutions have a bifurcate horizon, they put the left end of the
spacelike hypersurfaces at the bifurcation three-sphere, and they are
consistent with the constraints and preserved by the Hamiltonian evolution.
They also ensure that the cubic (\ref{y-cubic}) has a unique real solution
for $y$ near $r=0$. On a classical solution, the future unit normal vector
$n^a(t)$ to the spacelike hypersurfaces at $r=0$ then evolves according to
\begin{equation}
n^a(t_1) n_a(t_2) =
-\cosh\left(\int_{t_1}^{t_2}
\Lambda_0^{-1}(t) N_1(t) \, dt \right)
\ \ .
\label{n-boost}
\end{equation}

Next, consider the boundary conditions in the variational principle. At $r=0$,
we follow Refs.\ \cite{LW2,lau1,lou-win,BLPP} and make
$N_1 / \Lambda_0$ a prescribed function of~$t$. By~(\ref{n-boost}), this
means fixing the rate at which the constant $t$ hypersurfaces are boosted at
$r=0$. At $r=1$, we make $R$ and $-g_{tt} = N^2 -
{(\Lambda N^r)}^2$ prescribed positive-valued functions of~$t$. This
means fixing the intrinsic metric on the three-surface $r=1$, and in
particular fixing this metric to be timelike.

Finally, we need the boundary terms to be added to the bulk
action~(\ref{S-sigma}). As in
Ref.\ \cite{LW2}, it can be verified that the appropriate term at $r=0$ is
\begin{equation}
\int dt \, R_0 \left( \case{1}{3} R_0^2 + \lambdahat \right)
( N_1 / \Lambda_0 )
\ \ ,
\end{equation}
and the appropriate term at $r=1$ is the integral over $t$ of
\begin{eqnarray}
&&
N \Lambda ^{-1} R^2 R' - N^r \Lambda P_\Lambda
- \casehalf {\dot R} ( R^2 + \lambdahat)
\ln \left| {N + \Lambda N^r \over N - \Lambda N^r} \right|
\nonumber
\\
&&
+ \lambdahat N \left( {R' \over \Lambda} \right)
\left[
1 - {1 \over 3} {\left( {R' \over \Lambda} \right)}^2
- { {\dot R} ( {\dot R} - N^r R') \over N^2 }
\right]
\nonumber
\\
&&- {\lambdahat N^r \Lambda \over 3 N}
\left[
{{\dot R}^3 \over N^2 - {(\Lambda N^r)}^2 }
+
{N^r {(R')}^3 \over \Lambda^2}
\right]
\ \ .
\end{eqnarray}

We have therefore arrived at a variational principle with the desired
boundary conditions. The Lovelock generalization of Birkhoff's theorem
guarantees that classical solutions exist, and makes possible a
complete description of the solutions. One should note, however, that
our Hamiltonian action does not directly reflect the split of the
variables at $r=1$ into the dynamical degrees of freedom versus the
boundary data. With the data on the timelike
boundary, the evolution no longer forms a hyperbolic system.

\section{Canonical transformation and Hamiltonian reduction}
\label{sec:transformation}

In this section we simplify the constraints by a canonical
transformation and reduce the theory to unconstrained Hamiltonian
variables. The treatment will closely follow Refs.\
\cite{kuchar1,LW2,lou-win,BLPP}.

To begin, suppose that we are given the canonical data $( \Lambda, R,
P_\Lambda, P_R)$ on a spacelike hypersurface embedded in a classical
solution. We wish to reconstruct from this data both the spacetime and the
location of the hypersurface in the spacetime.


As we have noted above, the embedding of the hypersurface in the classical
solution defines a unambiguous value of~$y$: from the equation of
motion~(\ref{y-true}), one finds that this value is $y_{\rm true}:=
N^{-1}({\dot R} - N^r R')$. To reconstruct $y_{\rm true}$ from the canonical
data, one needs to solve the cubic~(\ref{y-cubic}), which may have up to
three real solutions. Near $r=0$,  the falloff (\ref{s-r}) guarantees that
the cubic has a unique real solution, and this solution must therefore be
equal to~$y_{\rm true}$. As $r$ increases, two spurious real solutions may
appear, but it is straightforward to verify that neither of the spurious real
solutions can ever be equal to~$y_{\rm true}$. Therefore, $y_{\rm true}$ is
recovered from (\ref{y-cubic}) by choosing the unique real root near $r=0$
and  following this root by continuity to all~$r$. We note that,
generically, neither of the spurious roots for $y$ satisfies the constraint
$H=0$.

After $y=y_{\rm true}$ has been recovered, the reconstruction proceeds in
full analogy with that in Ref.\ \cite{kuchar1}. The function $F$ appearing in
the metric (\ref{curv-metric}) is given by
\begin{equation}
F = { \left( {R' \over \Lambda}  \right) }^2 - y^2
\ \ ,
\label{F-def}
\end{equation}
and from (\ref{F-solution}) one finds for the mass the expression
\begin{equation}
M = \casehalf R^2 (1-F)
+ \case{1}{4} \lambdahat {(1-F)}^2
\ \ .
\end{equation}
Finally, one finds
\begin{equation}
T' = {\Lambda y \over F}
\ \ ,
\label{Tprime-def}
\end{equation}
which specifies the location of the hypersurface up to translations in the
Killing time. This completes the reconstruction.

Next, we wish to promote the reconstruction equations into a canonical
transformation, valid even when the equations of motion do not
hold. Provided we stay within a sufficiently narrow neighborhood of the
classical solutions, $y$ is again uniquely recovered as a function of
the canonical data by taking the unique real root of
(\ref{y-cubic}) near $r=0$ and continuously following this root as $r$
increases. Computing the Poisson bracket between $M$ and $T'$ suggests that
$-T'$ could serve as the momentum conjugate to~$M$; if this holds, the new
momentum conjugate to ${\sf R}:=R$ is fixed by the fact that the
supermomentum constraint generates spatial diffeomorphisms in all the
variables and must thus read $P_M M' + P_{\sf R} {\sf R}'$. These
considerations suggest the transformation
\begin{mathletters}
\label{trans}
\begin{eqnarray}
M &:=& \casehalf R^2 (1-F)
+ \case{1}{4} \lambdahat {(1-F)}^2 \ \ ,
\label{trans-M}
\\
P_M &:=& - {\Lambda y \over F}
\ \ ,
\label{trans-PM}
\\
{\sf R} &:=& R
\ \ ,
\label{trans-sfR}
\\
P_{\sf R} &:=&
F^{-1} \left( \Lambda^{-2} R' H_r - y H \right)
\ \ ,
\label{trans-PsfR}
\end{eqnarray}
\end{mathletters}%
with $F$
given by~(\ref{F-def}). We now need to examine whether
this transformation is indeed canonical.

To proceed, we arrange the difference of the integrands in the Liouville
forms as
\begin{eqnarray}
P_\Lambda \delta \Lambda
&&
+ P_R \delta R
- P_M \delta M
- P_{\sf R} \delta {\sf R}
\nonumber
\\
&&
=
\delta \!
\left[
\Lambda P_\Lambda - \lambdahat y \Lambda^{-1}{(R')}^2 +
\casehalf R' \left( R^2 + \lambdahat \right)
\ln \left| { R' + y\Lambda \over  R' - y\Lambda }
\right|^{\vphantom A}_{\vphantom A}
\right]
\nonumber
\\
&&\quad + \left\{
\left[ \lambdahat y \Lambda^{-1} R'
-
\casehalf \left( R^2 + \lambdahat \right)
\ln \left| { R' + y\Lambda \over  R' - y\Lambda }
\right|^{\vphantom A}_{\vphantom A}
\right]
\delta R
\right\}'
\ \ .
\label{diff-PdQ}
\end{eqnarray}
Both terms on the right-hand side of (\ref{diff-PdQ}) are well defined. Upon
integration from $r=0$ to $r=1$, the second term only produces contributions
from the two ends. The contribution from $r=0$ vanishes because of the
falloff~(\ref{s-r}). The contribution from
$r=1$ vanishes if $\delta R$ vanishes there. As $\delta$ should in
the context of the Liouville form be understood as a time derivative, this
happens when the boundary conditions fix $R$ to be independent of $t$ at
$r=1$. If this is the case, we see that the difference of the Liouville forms
is an exact form,
\begin{eqnarray}
&&\int_0^1 dr \left( P_\Lambda \delta \Lambda + P_R \delta R \right)
\ \
- \int_0^1 dr \left(P_M \delta M
+ P_{\sf R} \delta {\sf R} \right)
\nonumber
\\
&&
=
\delta
\left\{
\int_0^1 dr
\left[
\Lambda P_\Lambda - \lambdahat y \Lambda^{-1}{(R')}^2 +
\casehalf R' \left( R^2 + \lambdahat \right)
\ln \left| { R' + y\Lambda \over  R' - y\Lambda }
\right|^{\vphantom A}_{\vphantom A}
\right]
\right\}
\ \ ,
\label{int-diff-PdQ}
\end{eqnarray}
and the transformation is canonical.

If, on the other hand, the boundary conditions fix $R$ to be explicitly
$t$-dependent at $r=1$, one cannot similarly argue that $\delta R$ would
vanish at $r=1$.\footnote{This appears to have gone unmentioned
in Refs.\ \cite{LW2,lau1,BLPP}.}
As mentioned at the end of section~\ref{sec:metric}, the canonical variables
at $r=1$ do not cleanly split into ``independent" degrees of freedom versus
boundary data, and it us unclear to us what the proper attitude here should
be. We shall, nevertheless, proceed to regard the transformation as canonical
even when $R$ is explicitly $t$-dependent at $r=1$: as in Refs.\
\cite{LW2,lau1,BLPP}, it will be seen that no apparent inconsistency will
result. {}From the viewpoint of thermodynamics, the case of principal
interest will in any case be the one where $R$ is independent of $t$ at $r=1$.

By construction, our transformation is well defined in a sufficiently
narrow neighborhood of the classical solutions. It also has a unique
inverse. Equations (\ref{trans-M}) and~(\ref{trans-sfR}), together with the
falloff implied by~(\ref{s-r}), determine $F$ uniquely in terms of
$M$ and~${\sf R}$. Equations (\ref{F-def}) and (\ref{trans-PM}), together
with the fact that $\Lambda$ is by assumption positive, then determine
$\Lambda$ and~$y$. $P_\Lambda$~is obtained from~(\ref{y-cubic}), and $P_R$
finally from~(\ref{trans-PsfR}).

To obtain the action in the new variables, we note that the constraint terms
can be written as
\begin{equation}
NH + N^r H_r = N^M M' + N^{\sfR} P_\sfR
\ \ ,
\end{equation}
where
\begin{mathletters}
\begin{eqnarray}
N^M &=& - N F^{-1}\Lambda^{-1} R' - N^r F^{-1} \Lambda y
\ \ ,
\\
N^\sfR &=& Ny + N^r R'
\ \ .
\end{eqnarray}
\end{mathletters}%
This suggests that one could take $N^M$ and $N^\sfR$ as the
new independent Lagrange multipliers in the action. Examining
the falloff at $r=0$ reveals, however, that fixing $N^M$ at $r=0$ to a value
that is independent of the canonical variables is not equivalent to fixing
$N_1\Lambda_0^{-1}$ to a value that is independent of the canonical
variables. This difficulty can be remedied by redefining the Lagrange
multipliers near $r=0$ as in Ref.\ \cite{lou-win}, and the appropriate
boundary terms at $r=0$ and $r=1$ can then be constructed as in Refs.\
\cite{LW2,lou-win,BLPP}. After these steps, the constraints can be
eliminated by a Hamiltonian reduction as in Refs.\ \cite{kuchar1,LW2,BLPP},
and one recovers a reduced theory in a true Hamiltonian form.
The steps follow the cited references so closely that we shall here omit the
detail and proceed directly to the reduced action.

The reduced action reads
\begin{equation}
S_{\rm red} =
\int dt
\left(
\bp {\dot \bm}
- \bh \right)
\ \ .
\label{S-red}
\end{equation}
The coordinate $\bm$ arises from the unreduced theory as the $r$-independent
value that $M$ takes when the constraint $M'=0$ holds. The momentum
$\bp$ is related to the unreduced variables by
\begin{equation}
\bp := \int_0^1 dr \, P_M
\ \ .
\label{bfp}
\end{equation}
The Hamiltonian $\bh$ is given by
\begin{eqnarray}
\bh &=&
- {\sf N}_0
R_h \left( \case{1}{3} R_h^2 + \lambdahat \right)
\nonumber
\\
&&
- \left( B^2 + \lambdahat \right)
\left[
\sqrt{ Q^2 \sfF + {\dot B}^2}
+ \casehalf {\dot B}
\ln
\left(
{ \sqrt{ Q^2 \sfF + {\dot B}^2}  - {\dot B}
\over
\sqrt{ Q^2 \sfF + {\dot B}^2}  + {\dot B} }
\right)
\right]
\nonumber
\\
&&
+ \case{1}{3} \lambdahat Q^{-2}
{\left( Q^2 \sfF + {\dot B}^2 \right)}^{3/2}
\ \ ,
\label{bfh}
\end{eqnarray}
where
\begin{equation}
{\sf N}_0:=N_1/\Lambda_0
\ \ ,
\end{equation}
\begin{equation}
R_h := \sqrt{2\bm - \casehalf \lambdahat}
\ \ ,
\end{equation}
\begin{equation}
\sfF
:= 1 + {B^2 \over \lambdahat}
\left(
1 -
\sqrt{
1 + {4 \bm \lambdahat \over B^4}
}^{\vphantom A}
\right)
\ \ ,
\end{equation}
and $B$ and $Q^2$ are respectively the values of ${\sf R}$ and $-g_{tt}$ at
$r=1$. $B$, $Q^2$, and ${\sf N}_0$ are considered prescribed functions of~$t$,
satisfying $B>0$, $Q^2>0$, and ${\sf N}_0\ge0$. The range of $\bm$ is
$\case{1}{4}\lambdahat< \bm <\casehalf B^2 + \case{1}{4}\lambdahat$,
corresponding to $0<R_h<B$, and the range of $\bp$ is the full real axis.
$R_h$~is equal to~$R_0$, and on a classical solution it equals the horizon
radius.

The equation of motion for $\bm$ implies that $\bm$ is independent of~$t$:
the value of $\bm$ is simply the mass parameter of the classical solution.
The equation of motion for $\bp$ reflects the fact that,
by~(\ref{Tprime-def}), (\ref{trans-PM}), and~(\ref{bfp}), $\bp$ is equal to
the difference in the Killing times at the two ends of the spacelike
hypersurface.

\section{Quantization and the partition function}
\label{sec:quantization}

In this section we quantize the reduced Hamiltonian theory and obtain a
partition function as the trace of the analytically continued time
evolution operator.

{}From now on, we take the boundary radius independent of time,
${\dot B}=0$. We also subtract from the Hamiltonian (\ref{bfh}) the
value that the terms arising from $r=1$ would take on flat
spacetime. This subtraction does not affect the equations
of motion, but it does renormalize the value of the action: it is
analogous to subtracting the $K_0$ term in Einstein's
theory \cite{GH1,hawkingCC}. Writing $Q:=
\sqrt{Q^2} >0$, the new Hamiltonian is given by
\begin{equation}
\bh =
Q \! \left(1 - \sqrt{\sfF}\right)
\left[
B^2 + \case{1}{3} \lambdahat
\left(1 - \sqrt{\sfF}\right)
\left(2 + \sqrt{\sfF}\right)
\right]
- {\sf N}_0
R_h \left( \case{1}{3} R_h^2 + \lambdahat \right)
\ \ .
\label{bfh-static}
\end{equation}
The first of the two terms in (\ref{bfh-static}) is the Lovelock analogue of
the quasilocal energy of Brown and York \cite{BY-quasilocal,lau2,haw-hu}. The
second term arises from the bifurcation three-sphere, and it will give rise to
the black hole entropy.

Quantization proceeds exactly as in Refs.\ \cite{LW2,lou-win,BLPP}. We take
the wave functions $\psi$ to be functions of the configuration
variable~$\bm$, with $\case{1}{4}\lambdahat< \bm
<\casehalf B^2 + \case{1}{4}\lambdahat$, and we introduce an inner product
with some smooth and slowly varying weight factor. The Hamiltonian operator
is taken to act by multiplication by the function
$\bh$~(\ref{bfh-static}), $\psi(\bm) \mapsto \bh(\bm) \psi(\bm)$,
and the unitary time evolution operator is
easily found. We then analytically continue the arguments of the time
evolution operator to imaginary values: we set $\int Q dt = -i\beta$,
interpreting $\beta>0$ as the inverse temperature at the boundary, and $\int
{\sf N}_0 dt = -2\pi i$, motivated by the regularity of the classical
Euclidean solutions. The trace of the analytically continued time evolution
operator is divergent, but we can argue as in Refs.\
\cite{LW2,lou-win,BLPP} that an  acceptable renormalization is
achieved by introducing a suitable regularization, dividing by the trace of
the regularized identity operator, and finally eliminating the regulator. In
this fashion, we obtain for the renormalized trace the manifestly well-defined
expression
\begin{equation}
Z (\beta;B;\lambdahat)
=
{\left( \int_0^1 {\tilde \mu} dx \right)}^{-1}
\left[ \int_0^1 {\tilde \mu} dx \exp(-I_*) \right]
\label{Z}
\ \ ,
\label{partition}
\end{equation}
where the effective action $I_*$ is given by\footnote{This effective action
has been obtained previously \cite{simon-whiting} by the Euclidean
Hamiltonian reduction method of Ref.\ \cite{WYprl}.}
\begin{equation}
I_*
=
\beta B^2 \left( 1 - \sqrt{\sfF} \right) \left[ 1
+ {\lambdahat \over 3 B^2}
\left(1 - \sqrt{\sfF}\right)
\left(2 + \sqrt{\sfF}\right)
\right]
-
2\pi B^3 x
\left( \case{1}{3} x^2  + {\lambdahat \over B^2} \right)
\end{equation}
with
\begin{equation}
\sfF
=
1 + {B^2 \over \lambdahat}
\left( 1
- \sqrt{
1 + {2x^2 \lambdahat \over B^2} + {\lambdahat^2 \over B^4}
}
\right)
\ \ .
\label{sfF}
\end{equation}
We have introduced the integration variable $x=R_h/B$, and the smooth
and slowly varying positive function ${\tilde \mu}(x)$ arose from the choice
of the inner product.

We now interpret the object $Z(\beta;B;\lambdahat)$ (\ref{partition}) as
the partition function of a thermodynamical canonical ensemble describing
black holes in a spherical box with curvature radius $B$ and and inverse
boundary temperature~$\beta$. The thermodynamical properties of this ensemble
will be analyzed in the next section.

\section{Thermodynamics in the canonical ensemble}
\label{sec:thermodynamics}

As noted above, the partition function $Z(\beta;B;\lambdahat)$
(\ref{partition}) is manifestly well defined. Further, the form of the
integral in (\ref{partition}) guarantees that the (constant volume) heat
capacity, $C = \beta^2 \biglb(  \partial^2 (\ln Z) / \partial \beta^2
\bigrb)$, is always positive (see, for example, section IV of Ref.\
\cite{BLPP}), and that the ensemble has a well-defined density of states
\cite{york1,WYprl,whitingCQG,BWY1}. These properties support the
interpretation of the partition function in terms of a genuine
thermodynamical equilibrium ensemble, in spite of the fact that we arrived at
the partition function via an analytic continuation and not via direct
statistical mechanics arguments.

To proceed, we shall estimate the integral in (\ref{partition}) by the
saddle point approximation. We shall throughout assume  ${\tilde\mu}(x)$ to
be so slowly varying that its precise form will not affect the saddle point
analysis. We shall also assume that the action is sufficiently
rapidly varying to make the saddle point approximation is justified, without
attempting to explicitly state the necessary conditions; typically, it will
be throughout assumed that the system is ``macroscopic," $B\gg 1$.

The critical points of $I_*$ are at the roots of the equation
\begin{equation}
{\beta x \over 2 \pi B}
=
\left( x^2 + {\lambdahat \over B^2 } \right)
\sqrt{\sfF}
\ \ .
\label{critpointseq}
\end{equation}
The critical points give precisely the Lorentzian black hole solutions whose
Hawking temperature at the boundary, calculated in the usual way from the
surface gravity \cite{myers-simon} and the blueshift factor, is equal
to~$\beta$. The mass of the hole is $\bm = \case{1}{2}B^2 x^2 +
\case{1}{4}\lambdahat$, and the value of $I_*$ at a critical point equals the
Euclidean action of the corresponding Euclideanized black hole solution.
Whenever the partition function is dominated by a critical point, we recover
for the thermal energy expectation value and the entropy the results
\begin{mathletters}
\label{e-and-s-gen}
\begin{eqnarray}
\langle E \rangle
&=&
 - {\partial (\ln Z) \over \partial \beta}
\approx
B^2 \left( 1 - \sqrt{\sfF} \right) \left[ 1
+ {\lambdahat \over 3 B^2}
\left(1 - \sqrt{\sfF} \right)
\left(2 + \sqrt{\sfF} \right)
\right]
\ \ ,
\label{E-expect}
\\
S
&=&
\left( 1 - \beta {\partial \over
\partial\beta}\right) (\ln Z)
\approx
2\pi B^3 x
\left( \case{1}{3} x^2  + {\lambdahat \over B^2} \right)
=
2\pi R_h
\left( \case{1}{3} R_h^2  + \lambdahat \right)
\ \ ,
\label{entropy}
\end{eqnarray}
\end{mathletters}%
where $x$ and $\sfF$ are evaluated at the critical point. The
expression (\ref{entropy}) for the entropy agrees with the result first
obtained by Euclidean methods \cite{myers-simon}.

We can now extract physical information by analyzing the critical point
structure of $I_*$ in various limits of interest in the three parameters
$\lambdahat$, $B$, and~$\beta$.

As a preliminary, consider the case $\lambdahat=0$, in which our Lovelock
theory reduces to Einstein's theory. Although we have for presentational
simplicity assumed $\lambdahat>0$, it is easy to see that the partition
function for Einstein's theory is correctly recovered by taking the limit
$\lambdahat\to0$ in equations (\ref{partition})--(\ref{sfF}). In particular,
(\ref{sfF}) reduces to $\sfF = 1 - x^2$, and the critical point equation
(\ref{critpointseq}) reduces to
\begin{equation}
{\beta \over 2 \pi B}
=
x
\sqrt{1 - x^2}
\ \ .
\label{einstein-critpointseq}
\end{equation}
The condition for critical points to exist is $\beta \le \pi B$, and the
critical points are then at $x = x_\pm := 2^{-1/2}{\left( 1 \pm \sqrt{1 -
\pi^{-2} B^{-2} \beta^2} \right)}^{1/2}$. When the critical points are
distinct, $x_+$ is a local minimum and $x_-$ a local
maximum. When
$\beta^2 < \case{3}{4} \pi^2 B^2$, the partition function gets its
dominant contribution from the global minimum at $x=x_+$.  When $\beta^2 >
\case{3}{4} \pi^2 B^2$, on the other hand, the partition function gets its
dominant contribution from the vicinity of the global minimum at $x=0$. The
limiting case $\beta^2 = \case{3}{4} \pi^2 B^2$ represents a phase transition
where the dominant contribution shifts from $x=x_+$ to $x=0$ as
$\beta$ increases. When the saddle point dominates, the thermal energy and
entropy (\ref{e-and-s-gen}) take the form
\begin{mathletters}
\label{e-and-s-einstein}
\begin{eqnarray}
\langle E \rangle &\approx& B^2 \left( 1 - \sqrt{1-x_+^2} \right)
\ \ ,
\label{einstein-energy}
\\
S
&\approx&
\case{2}{3}\pi B^3 x_+^3 = \case{2}{3}\pi R_h^3
\ \ ,
\label{einstein-entropy}
\end{eqnarray}
\end{mathletters}%
and the relation between the thermal energy and the
mass can be written as
\begin{equation}
\bm \approx \langle E \rangle - {{\langle E \rangle}^2 \over 2 B^2}
\ \ .
\label{mass-energy}
\end{equation}
Equation (\ref{mass-energy}) displays explicitly how the mass gets a
contribution both from the thermal energy and from the gravitational binding
energy associated with the thermal energy. Expectedly, the situation is
closely similar to that in four-dimensional Einstein theory
\cite{york1,WYprl,BWY1}.

We now turn to the case $\lambdahat>0$, in which $I_*$ always has at least
one critical point.

Consider first the limit of small $\lambdahat$ with fixed $B$ and~$\beta$.
The situation differs from that in the case $\lambdahat=0$ only in that there
is now one new critical point, a local minimum, at $x = 2\pi \lambdahat
B^{-1}\beta^{-1} + O(\lambdahat^2)$. At the new critical point, $I_* =
\case{1}{4} \lambdahat\beta + O(\lambdahat^2)$. Therefore, as
$\lambdahat\to0$, the partition function smoothly approaches that of
Einstein's theory. In particular, when $\beta^2 < \case{3}{4} \pi^2 B^2$, it
would be straightforward to compute the first order correction in
$\lambdahat$ to the thermal energy and the entropy~(\ref{e-and-s-einstein}),
assuming that the corrections to the saddle point approximation are small.

Consider next the small $\beta$ limit with fixed $B$ and~$\lambdahat$. There
is only one critical point, at $x = 1 - \case{1}{8}\pi^{-2}{(B^2 +
\lambdahat)}^{-1}
\beta^2 + O(\beta^4)$, and this critical point is the global minimum
of~$I_*$. One can think of this critical point as the counterpart of the
larger of the two critical points of the case $\lambdahat=0$: the black hole
fills almost all of the box. The disappearance of the smaller critical point
of the case $\lambdahat=0$ is related to the fact that, for
fixed~$\lambdahat$, the Hawking temperature of the Lovelock hole in
asymptotically flat space is bounded below by $\case{1}{4} \pi^{-1}
\lambdahat^{-1/2}$ \cite{myers-simon}. If the saddle point approximation to
the partition function remains good, the thermal energy and the entropy are
given by
\begin{mathletters}
\label{e-and-s-smallbeta}
\begin{eqnarray}
\langle E \rangle
&\approx&
B^2 +
\case{2}{3}\lambdahat - \casehalf \pi^{-1}B
\beta + O(\beta^2)
\ \ ,
\\
S
&\approx&
2\pi B^3
\left(
{1\over3} + {\lambdahat \over B^2}
- {\beta^2 \over 8 \pi^2 B^2}
\right)
+ O(\beta^4)
\ \ .
\end{eqnarray}
\end{mathletters}%

Consider next the large $\beta$ limit with fixed $B$ and~$\lambdahat$. There
is again only one critical point, at $x = 2\pi \lambdahat B^{-1} \beta^{-1}
\sfF_0^{1/2} + O(\beta^{-3})$, where $\sfF_0 := 1 + B^2
\lambdahat^{-1} \left( 1 - \sqrt{1 + \lambdahat^2 B^{-4}} \right)$.
This critical point is the global minimum, and it has no counterpart in
Einstein's theory: it corresponds to a small, ``purely Lovelock," black
hole. If the saddle point approximation to the partition function remains
good, the thermal energy and the entropy are easily read off
from (\ref{e-and-s-gen}) as
\begin{mathletters}
\label{e-and-s-largebeta}
\begin{eqnarray}
\langle E \rangle
&\approx&
B^2 \left( 1 - \sfF_0^{1/2} \right) \left[ 1
+ {\lambdahat \over 3 B^2}
\left(1 - \sfF_0^{1/2}\right)
\left(2 + \sfF_0^{1/2}\right)
\right]
+ O(\beta^{-2})
\ \ ,
\\
S
&\approx&
{4\pi^2 \lambdahat^2 \sfF_0^{1/2}
\over
\beta}
+ O(\beta^{-3})
\ \ .
\end{eqnarray}
\end{mathletters}%

Consider then the large $\lambdahat$ limit with fixed $B$ and~$\beta$. There
is again only one critical point, at $x = 1 - \case{1}{8} \pi^{-2} \beta^2
\lambdahat^{-1} + O(\lambdahat^{-2})$, and this critical point is the global
minimum. The hole is again ``purely Lovelock," but it now fills almost all of
the box.

Finally, consider the large $B$ limit with fixed $\lambdahat$ and~$\beta$.
One critical point is at $x = 1 - \case{1}{8} \pi^{-2} B^{-2} \beta^2 +
O(B^{-4})$. This critical point is the global minimum, and it can be regarded
as the counterpart of the larger of the two critical points of the case
$\lambdahat=0$. If $\beta> 4
\pi \lambdahat^{1/2}$, there are in addition two other critical points, at
$x = \case{1}{4} \pi^{-1} B^{-1} \beta \left( 1 \pm \sqrt{1 -
16 \pi^2 \lambdahat \beta^{-2}} \right) + O(B^{-3})$. The fact that the two
small critical points exist only for $\beta> 4 \pi \lambdahat^{1/2}$ is
related to the above-mentioned phenomenon that the  Hawking temperature of
our Lovelock hole in asymptotically flat space is bounded below by
$\case{1}{4} \pi^{-1} \lambdahat^{-1/2}$ \cite{myers-simon}. If the saddle
point approximation is good, the thermal energy and the entropy are obtained
by replacing both $O$-terms in (\ref{e-and-s-smallbeta}) by~$O(B^{-1})$.

We therefore see that for $\lambdahat>0$, the partition function is always
dominated by a black hole solution in the limits that we have considered. In
the high temperature limit and in the large box limit, the situation is
very similar to that for $\lambdahat=0$, in that the dominating
black hole solution fills most of the box. In the low temperature limit, on
the other hand, the Lovelock theory does exhibit a dominating black hole
solution where none existed in the case $\lambdahat=0$. For a macroscopic box
and $\lambdahat\ll B^2$, however, the presence of the new dominating solution
does not appear to make the thermodynamical behavior qualitatively
different from that in the case $\lambdahat=0$. One can read these results
as evidence for stability of the qualitative thermodynamical behavior of
Einstein's theory upon the addition of the Lovelock parameter.

For $\lambdahat >0$, the critical point structure of $I_*$ is entirely
determined by two parameters, which can be conveniently taken to be
$\lambdahat B^{-2}$ and~$\beta B^{-1}$. Numerical experimentation suggests
that there are never more than three critical points. When
$\lambdahat B^{-2}$ is fixed and sufficiently large, there is only one
critical point for any~$\beta B^{-1}$: this critical point is the global
minimum, and it migrates smoothly from the large Lovelock hole to the small
Lovelock hole as $\beta B^{-1}$ increases. When $\lambdahat B^{-2}$ is fixed
and sufficiently small, on the other hand, the transition between the unique
minima for small and large $\beta B^{-1}$ takes place via a phase in which
$I_*$ has three critical points, a maximum surrounded by two minima: as
$\beta B^{-1}$ tends to zero (infinity), only the larger (smaller) of the
two minima prevails. When three critical points exist, one can find
regions of the parameter space where the global minimum is at either of the
two local minima. We have, however, not attempted to corroborate these
numerical experiments analytically.

It should be emphasized that the partition function has no well-defined limit
as $B\to\infty$ with fixed $\beta$ and~$\lambdahat$, neither for
$\lambdahat=0$ nor for $\lambdahat>0$. As with Einstein's theory in four
dimensions \cite{york1,WYprl}, this reflects the fact that the thermodynamical
canonical ensemble is not well defined in asymptotically flat space. We shall
give a more detailed comparison of the boxed Lovelock theory to Lovelock
theory in asymptotically flat space in appendix~\ref{app:as-flat}\null.

\section{Summary and discussion}
\label{sec:discussion}

In this paper we have investigated the Hamiltonian dynamics and thermodynamics
of five-dimensional spherically symmetric Lovelock theories in which the only
contributions to the Lagrangian density are the Einstein-Hilbert term and the
four-dimensional Euler density. We adopted boundary conditions that enforce
every classical solution to be part of the exterior region of a black hole,
with the spacelike  hypersurfaces extending from the horizon bifurcation
three-sphere to a timelike boundary with fixed intrinsic metric. We
simplified the constraints by a canonical transformation that generalizes the
one introduced by Kucha\v{r} in four-dimensional spherically symmetric
Einstein theory, and we reduced the theory classically to its true dynamical
degrees of freedom.

After Hamiltonian quantization, we interpreted the trace
of the analytically continued time evolution operator as the partition
function of a thermodynamical canonical ensemble, describing
black holes in a spherical box whose size and boundary temperature are fixed.
In the special case where the Lovelock parameter $\lambda$ vanishes and the
theory reduces to Einstein's theory, we found that the thermodynamics is
highly similar to that of the corresponding system in
four-dimensional Einstein theory: in particular, for high boundary
temperatures the partition function is dominated by a classical black hole
solution that fills most of the box. When $\lambda>0$, the
situation was more versatile. In the high temperature limit, with $\lambda$
and the box size fixed, the partition function is again dominated by a black
hole that fills most of the box. In the low temperature limit, on the other
hand, the partition function is now {\em also\/} dominated by a black
hole solution; this black hole is small, and it has has no analogue in
Einstein theory. Nevertheless, if $\lambda$ is small compared with the size
of the box, the new dominating solution has little qualitative effect on the
thermodynamical properties. In this sense, the qualitative thermodynamical
behavior of the Einstein system is stable upon the addition of the Lovelock
parameter.

When the box size is taken to infinity, we found that the partition function
has no well-defined limit, neither for
$\lambda=0$ nor for $\lambda>0$. While this is not surprising for
Einstein's theory, in view of the similar phenomenon in four dimensions
\cite{york1,WYprl}, one might have hoped the theory with $\lambda>0$ to
fare better on the grounds that this theory admits asymptotically flat
black hole solutions that are stable against Hawking evaporation
\cite{myers-simon}. However, even though a classical solution that dominates
a well-defined partition function must be stable against Hawking evaporation
\cite{whitingCQG}, our Lovelock theory in asymptotically flat space provides
an example where the mere existence of such a locally stable classical
solution does not imply the existence of well-defined canonical ensemble.
Another such example occurs in four-dimensional Einstein-Maxwell theory in
asymptotically flat space \cite{lou-win}.

In the classical theory with $\lambda>0$, we saw that the super-Hamiltonian
emerges as a multivalued function of the canonical variables, as is
generically the case in Lovelock theories
\cite{teit-zan-love1,teit-zan-love2}. Nevertheless, our thermodynamically
motivated boundary conditions were sufficient to uniquely specify the
super-Hamiltonian near the horizon, and the uniqueness could be extended to
the full spacelike hypersurfaces by continuity. Another boundary condition
that would uniquely specify the super-Hamiltonian in this fashion is the
asymptotically flat falloff (\ref{l-r}) discussed in
appendix~\ref{app:as-flat}. However, one expects there to exist boundary
conditions of interest for which such uniqueness does not occur, and in such
cases one would need to seek other criteria for specifying the
super-Hamiltonian. If one regards the Lovelock theory as a perturbation to
Einstein's theory, or as a toy model for semiclassical gravity with
back-reaction, one possible criterion of this kind is perturbative
expandability of the solutions in~$\lambda$ \cite{jzs-pertur}.

In conclusion, our results provide evidence for robustness of the classical
Hamiltonian structure and the qualitative thermodynamical structure of
spherically symmetric Einstein gravity in five dimensions upon the addition
of the four-dimensional Euler density in the action. To put this conclusion in
proper perspective, one should remember that both our particular Lovelock
theory and our boundary conditions were hand-picked so that the global
aspects of the problem remained virtually identical to those in pure Einstein
gravity. It is tempting to think that this may exemplify a more general
connection between the global properties of the (space of) classical
solutions and the qualitative behavior of thermodynamical equilibrium
ensembles: one might conjecture that whenever the global properties of a
Lovelock theory are sufficiently similar to those of Einstein's theory, then
also the equilibrium thermodynamics, with finite or infinite boundary
conditions, will be qualitatively similar to that in Einstein's theory.
Another example supporting such a conjecture is provided by the
asymptotically anti-de~Sitter Lovelock theories of Ref.\ \cite{btz-cont},
which include as a special case Einstein gravity in three and four dimensions
with a negative cosmological constant.  However, to give the conjecture a
more substantial meaning, one would need a more systematic understanding of
the possible global structures that the various Lovelock theories may have.

\acknowledgments
We would like to thank John Friedman, Eric Poisson, and Bernard Whiting for
helpful discussions. Some of the results in sections \ref{sec:quantization}
and \ref{sec:thermodynamics} overlap with previous work by one of us (J.Z.S)
and Bernard Whiting \cite{simon-whiting}. Much of the work was done using
MATHEMATICA \cite{mathematica} and MATHTENSOR \cite{mathtensor}: in
particular, the spherically symmetric action (\ref{S-lag-sigma}) was derived
independently both on a computer and by hand, and we used a computer to
verify that the equations of motion derived from (\ref{S-lag-sigma}) are
equivalent to the correct spherically symmetric equations, derived
from~(\ref{love-action-gen}).  This work was supported in part by NSF grants
PHY-94-13253,
PHY-94-21849,
and
PHY-95-07740.

\appendix
\section{Lovelock black hole}
\label{app:love-hole}

In this appendix we briefly review the classical solutions to our Lovelock
theory.

By the Lovelock generalization of Birkhoff's theorem \cite{wiltshire1}, the
general solution to the theory (\ref{love-action-gen}) for $\lambda\ne0$ can
be written in the local curvature coordinates $(T,R)$ as
\begin{equation}
ds^2 = - F dT^2 + F^{-1} dR^2 + R^2 d\Omega_3^2
\ \ ,
\label{curv-metric}
\end{equation}
where
\begin{equation}
F = 1 + {R^2 \over \lambdahat}
\left(
1 \pm
\sqrt{
1 + {2 \omega \lambdahat \over R^4}
}^{\vphantom A}
\right)
\label{F-solution}
\end{equation}
with $\lambda = \casehalf\lambdahat$. The coordinates $T$ and $R$
are respectively the Killing time, whose constant value hypersurfaces are
spacelike or timelike depending on the sign of~$F$, and the curvature radius
of the three-sphere. For $\lambda=0$, the general solution is the
five-dimensional Schwarzschild solution, obtained from (\ref{F-solution})
with the lower sign as the limit $\lambdahat\to0$. In addition to~$\lambda$,
the only local parameter in the solution is the mass-like quantity~$\omega$.

For the reasons discussed in section~\ref{sec:intro}, we take
$\lambdahat\ge0$. The solution then describes a black hole in
asymptotically flat space provided we take $\omega>
\casehalf \lambdahat$ and, for $\lambdahat>0$, choose the lower sign
in (\ref{F-solution}) \cite{myers-simon}. The curvature coordinates are
good individually in each region not containing horizons: the horizons are
nondegenerate, and the Penrose diagram of the conventional maximal analytic
extension is similar to that of Kruskal manifold. The horizon radius is $R_h
= \sqrt{\omega - \casehalf \lambdahat}$. In our units, the ADM mass
is $M =\casehalf \omega$.

\section{Asymptotically flat infinity}
\label{app:as-flat}

In this appendix we adapt the analysis of the main text to boundary
conditions that replace the timelike boundary by an
asymptotically flat infinity. We shall see that quantization along the lines
of section \ref{sec:quantization} will not lead to a well-defined
canonical ensemble.

In the metric theory of section~\ref{sec:metric}, we let $r$ take the range
$0\le r < \infty$. At $r\to\infty$, we introduce the falloff
\begin{mathletters}
\label{l-r}
\begin{eqnarray}
\Lambda(t,r)
&=&
1 + M_+(t) r^{-2} + O^\infty \left( r^{-2-\epsilon} \right)
\ \ ,
\label{l-r-Lambda}
\\
R(t,r)
&=&
r + O^\infty \left( r^{-1-\epsilon} \right)
\ \ ,
\label{l-r-R}
\\
P_{\Lambda}(t,r) &=& O^\infty \left( r^{1-\epsilon} \right)
\ \ ,
\label{l-r-PLambda}
\\
P_{R}(t,r) &=& O^\infty \left( r^{-\epsilon} \right)
\ \ ,
\label{l-r-PR}
\\
N(t,r) &=& N_+(t) + O^\infty \left( r^{-\epsilon} \right)
\ \ ,
\label{l-r-N}
\\
N^r(t,r) &=& O^\infty \left( r^{-1-\epsilon} \right)
\ \ ,
\label{l-r-Nr}
\end{eqnarray}
\end{mathletters}%
where $N_+(t)>0$, and $\epsilon$ is a parameter that can be chosen
arbitrarily in the range $0<\epsilon\leq2$. $O^\infty(r^{s})$ denotes a term
that is bounded at $r\to\infty$ by a constant times~$r^{s}$, and whose
derivatives fall off accordingly. It is straightforward to verify that this
falloff makes the coordinates asymptotic to hyperspherical coordinates in
Minkowski space, it is consistent with the constraints, and it is
preserved by the time evolution. $N_+(t)$ gives the rate at which
the asymptotic Minkowski time advances with respect to the coordinate
time~$t$. When the equations of motion hold, $M_+(t)$ is independent of~$t$,
and its value is the ADM mass.

The total action takes the form $S = S_\Sigma + S_{\partial\Sigma}$,
where
\begin{equation}
S_{\partial \Sigma}
=
\int dt \, R_0 \left( \case{1}{3} R_0^2 + \lambdahat \right)
( N_1 / \Lambda_0 )
- \int dt \, N_+ M_+
\ \ ,
\end{equation}
and $S_\Sigma$ is as in (\ref{S-sigma}) except that the upper limit of the
$r$-integral is replaced by infinity. The canonical transformation and
Hamiltonian reduction proceed as in section~\ref{sec:transformation}. The
action of the reduced theory is as in~(\ref{S-red}), but with the
Hamiltonian now given by
\begin{equation}
\bh =
N_+ \bm
- {\sf N}_0
R_h \left( \case{1}{3} R_h^2 + \lambdahat \right)
\ \ .
\label{bfh-flat}
\end{equation}
The configuration variable $\bm$ has the range $\bm >
\case{1}{4}\lambdahat$, while the range of $\bp$ is the full real axis.
Although in the main text we assumed for presentational simplicity
$\lambdahat>0$, it is easily verified that the above derivation
extends to the case $\lambdahat=0$, and the reduced Hamiltonian
(\ref{bfh-flat}) is valid for all $\lambdahat\ge0$.

Quantization can now proceed as in the main text.
We analytically continue the time evolution operator by
$\int {\sf N}_0 dt = -2\pi i$ and $\int N_+ dt = -i\beta$, interpreting
$\beta$ as the inverse temperature at the infinity. For the renormalized
trace of the analytically continued time evolution operator, we obtain
formally
\begin{equation}
Z (\beta;\lambdahat)
=
{\cal N}
\int_0^\infty {\tilde \mu} dR_h \exp(-I^\infty_*)
\label{Z-flat}
\ \ ,
\label{partition-flat}
\end{equation}
where the effective action $I^\infty_*$ is given by
\begin{equation}
I^\infty_*
=
\casehalf \beta \left( R_h^2  + \casehalf \lambdahat \right)
-
2\pi R_h
\left( \case{1}{3} R_h^2  + \lambdahat \right)
\ \ .
\end{equation}
The smooth and slowly varying positive function ${\tilde \mu}(R_h)$ arose
from the choice of the inner product, and ${\cal N}$ is a normalization
constant, possibly dependent on $\lambdahat$ but presumably not on~$\beta$.
However, the integral in
(\ref{partition-flat}) is divergent because $I^\infty_*$ tends to $-\infty$
at large~$R_h$. Thus, the canonical ensemble does not exist under the
asymptotically flat boundary conditions, neither for $\lambdahat=0$ nor
$\lambdahat>0$. In this sense, the asymptotically flat Lovelock theory is
thermodynamically no better behaved than asymptotically flat Einstein
theory.

The critical points of $I^\infty_*$ give the (Lorentzian) classical solutions
that have the inverse Hawking temperature $\beta$ at infinity. For
$\lambdahat=0$ there exists exactly one critical point, which is a local
maximum: this is similar to what happens with Einstein's theory in four
dimensions
\cite{york1}. For $\lambdahat>0$, the situation is more versatile.
Critical points exist when $\beta^2 \ge 16 \pi^2 \lambdahat$,
and when the inequality is genuine, there are two critical points. The
critical point with the smaller (larger) value of $R_h$ is a local minimum
(maximum, respectively). The local minimum gives the classical solution
that was found to be stable against Hawking evaporation in Ref.\
\cite{myers-simon}. While the stability of this solution against Hawking
evaporation reflects its being a local minimum of~$I^\infty_*$
\cite{whitingCQG}, the divergence of the integral in (\ref{Z-flat})
demonstrates that this local stability is not sufficient to guarantee the
existence of the canonical ensemble. The effects of the Lovelock parameter on
the asymptotically flat thermodynamics are thus qualitatively very similar to
those of a fixed charge in asymptotically flat four-dimensional
Einstein-Maxwell theory \cite{lou-win,davies1,davies2}.

\end{document}